# HOM AND FP COUPLER DESIGN FOR THE NLSF HIGH GRADIENT SC CAVITY

N. Juntong and R.M. Jones
The Cockcroft Institute, Daresbury, Warrington, Cheshire WA4 4AD, UK.
The University of Manchester, Oxford Road, Manchester, M13 9PL, UK.

*Abstract*

The design of both higher order mode (HOM) and fundamental power (FP) couplers for a unique New Low Surface Field (NLSF) cavity [1] is presented. Here we present a study which uses ILC baseline couplers. The Balleyguier method [2] of calculating external quality factor is used and the results validated using both Microwave studio and HFSS.

## INTRODUCTION

The FP coupler for an SRF cavity can either be waveguide or coaxial in nature. Both have some advantages and disadvantages in term of mechanical design, power handling, and multipacting. A choice of coupler is made on the basis of operating frequency, source power, and simplicity of design. At lower frequencies the waveguide is quite large and it is cumbersome to accommodate it in the cryostat, while a coaxial type will be compact. On the other hand because of its large size, external cooling is readily provided by brazing on cooling channels on the waveguide coupler design. A coaxial type will be used in the ILC because of its compact size and it can easily be fitted inside the cryomodule. Furthermore, modifying the external quality factor ($Q_e$) is more straightforward.

The FP coupler transfers rf power from the source system to the cavity. This is achieved by providing an impedance match between the source system and the combined cavity-beam system. For heavy beam loading in an SRF cavity case, the optimal $Q_e$ for operating with zero reflection is [3]:

$$Q_e = \frac{V_c^2}{P_b(R/Q_0)} \quad (1)$$

where $V_c$ is cavity voltage, $P_b$ is beam power, R is shunt impedance, and $Q_0$ is the cavity fundamental mode quality factor. For the ILC, the average $Q_e$ is prescribed to be $3.5 \times 10^6$ [4].

When intense charged particle bunches travel through cavities they excite electromagnetic (e.m.) waves at a host of frequencies. If left undamped, these beam-excited HOMs can dilute the emittance of the beam and, in the worst case scenario, can also cause a beam breakup (BBU) instability to occur. In practise the HOMs are damped down to a level prescribed by beam dynamics simulations. The $Q_e$s of both the FP and HOM coupler are simulated and the sensitivity to axial and angular positioning is assessed.

This paper is organised such that the method used to obtain the $Q_e$ is described in the next section. The following section presents the validation of this method on the NLSF cavity with both finite element and finite difference electromagnetic solver codes. Thereafter it is applied to design the FP and HOM couplers for the NLSF cavity. Some concluding remarks are presented in the final section.

## BALLEYGUIER METHOD

This was developed to obtain the $Q_e$ of a cavity-coupler system [2]. It relies on combining two standing wave solutions to represent a travelling wave in the coupler line. The method starts by considering a lossless cavity with stored rf energy, U, at resonant frequency, ω. The cavity power lost into the coupler line is P. The $Q_e$ (=ωU/P) can be expressed as

$$Q_e = \frac{\omega \iiint |F|^2 dV}{c \iint |F|^2 dS} \quad (2)$$

for a single TEM mode propagating in a coupler line under vacuum, where F is either the electric or magnetic field.

The method utilises two terminated boundary conditions at the coupler line: electric and magnetic. With a magnetic termination an electric field will have maximum amplitude on the plane, which is larger than the individual waves. From Eq. 2 we can define the $Q_1$ as

$$Q_1 = \frac{\omega \iiint_{\text{cavity}} |E_1|^2 dV}{c \iint_{\text{ref.plane}} |E_1|^2 dS} = \frac{|1+e^{j\varphi}|^2}{4} Q_e \quad (3)$$

with φ is the phase difference between the two fields. In a similar fashion for an electric termination, here a magnetic field reaches a maximum and we define $Q_2$ as

$$Q_2 = \frac{\omega \iiint_{\text{cavity}} |H_2|^2 dV}{c \iint_{\text{ref.plane}} |H_2|^2 dS} = \frac{|1-e^{j\varphi}|^2}{4} Q_e \quad (4)$$

Combining these two allows $Q_e$ to be obtained as

$$Q_e = Q_1 + Q_2 \quad (5)$$

This is independent of φ since: $|1+e^{j\varphi}|^2 + |1-e^{j\varphi}|^2 = 4$. Thus two eigenmode simulation runs are adequate to obtain the $Q_e$ of the system.

## VALIDATION OF METHOD

To validate this technique, it is applied to a single-cell NLSF cavity equipped with a FP coaxial coupler as illustrated in Fig. 1. In this configuration the coaxial coupler is located 45 mm away from the entrance to the cavity. The antenna tip is located at the same vertical level as the iris. Two simulation have been carried out with different terminating boundary conditions in both MWS [5] and HFSS [6]. Additionally, one eigensolution is performed using MWS with the $Q_e$ calculation module selected. The simulation results are compared in Table 1.

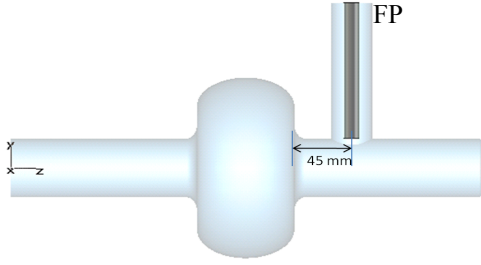

Figure 1: Single-cell NLSF FP for a cavity-coupler validation simulation.

Table 1: Comparison of the $Q_e$ produced using MWS and HFSS on a single-cell NLSF cavity.

| Parameter | $Q_1$ [x$10^6$] | $Q_2$ [x$10^6$] | $Q_e$ [x$10^6$] | $\omega/2\pi$ [GHz] |
|---|---|---|---|---|
| HFSS (A) | 3.28 | 1.18 | 4.46 | 1.289 |
| MWS (B) | 3.38 | 1.13 | 4.51 | 1.287 |
| MWS built-in (C) | - | - | 4.66 | 1.287 |
| |A-B|/A [%] | 3.0 | 4.2 | 1.1 | - |
| |A-C|/A [%] | - | - | 4.5 | - |

There is reasonable agreement between the independent codes on the $Q_e$s. In order to allow these results to converge a large mesh was necessary. 1.2x$10^6$ tetrahedra were used in HFSS and 1.0x$10^6$ mesh cells in MWS. The maximum discrepancy between the 3 methods is less than 5%.

This method has also been applied to a full nine-cell NLSF cavity with the same standard coaxial coupler configuration as the single-cell cavity-coupler system.

Table 2: Balleyguier $Q_e$ calculation on a nine-cell NLSF cavity attaches to FP couplers.

| Parameter | $Q_1$ [x$10^6$] | $Q_2$ [x$10^6$] | $Q_e$ [x$10^6$] | $\omega/2\pi$ [GHz] |
|---|---|---|---|---|
| HFSS (A) | 2.95 | 0.88 | 3.83 | 1.300 |
| MWS (B) | 2.76 | 0.91 | 3.86 | 1.296 |
| MWS built-in (C) | - | - | 3.77 | 1.296 |
| |A-B|/A [%] | 6.4 | 3.4 | 0.8 | - |
| |A-C|/A [%] | - | - | 1.6 | - |

The mode frequencies and $Q_e$s are listed in Table 2. Here the maximum discrepancy is no more than ~6%. These results provide some confidence on the accuracy and applicability of the method. We then applied this technique to several coupler configurations. This is discussed in the following sections.

## FUNDAMENTAL POWER COUPLER

Various shapes of coaxial couplers are studied with a view to optimising the coupling. All couplers are parameterised as indicated in Fig. 2.

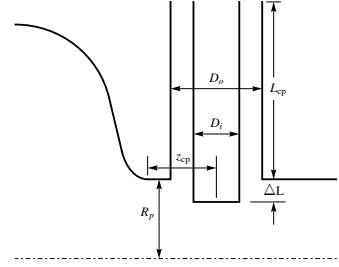

Figure 2: Schematic of the FP coupler

We aim at achieving $Q_e \sim 3.5$ x$10^6$, similar to the ILC. This ensures that the perturbation on the accelerating field flatness is minimised. In all simulations, the eigenmode module of MWS was utilised to obtain $Q_e$ and these results are displayed in Fig. 3. Out of several possible couplers, we opt for the standard TTF-III coupler [7] for our NLSF cavity. In order to obtain the requisite Q, the coupler is placed 45 mm away from cavity entrance and has an antenna penetration depth of 6 mm. This configuration also minimises the perturbation on the accelerating field. Other designs may enhance the multipacting in this region and hence are less favoured. The HOM couplers are considered in the next section.

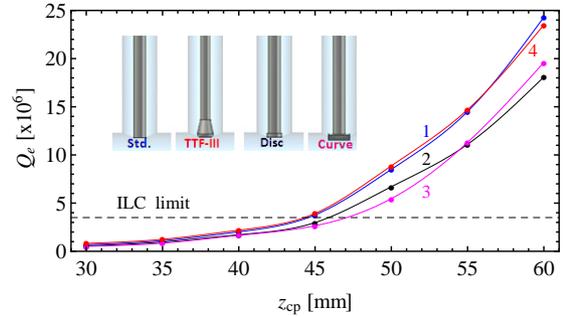

Figure 3: $Q_e$ vs coupler position for various couplers: a simple coaxial coupler (1-indicated by blue dots), a disc-type coupler (2-indicted by black dots), a curve-type coupler (3-indicted by magenta dots), and a TTF-III coupler (4-indicted by red dots), where the ILC requirement is indicted by a dashed line.

## HOM COUPLER

In order to damp unwanted beam-excited HOMs, which can degrade beam characteristics, HOM couplers need to be carefully designed. Here, the baseline TESLA HOM coupler is studied. This is shown inset in Fig. 4.

Firstly, we investigated using the same configuration as used in the TESLA design, i.e. HOM couplers were placed 115 degrees with respect to each other, as indicated in Fig. 4.

Simulations were performed to obtain the $Q_e$s of the first three dipole bands using both MWS and HFSS. These three bands have significantly larger R/Qs than the other bands [1].

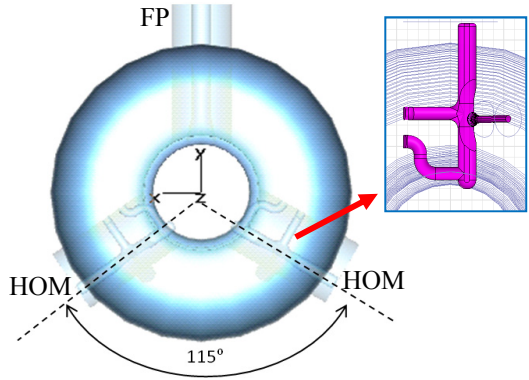

Figure 4: Baseline TESLA HOM couplers configuration in the NLSF cavity. HOM coupler antenna simulation model is shown inset.

In HFSS simulations $2 \times 10^6$ tetrahedra were used and $1.6 \times 10^6$ mesh cells in MWS. The $Q_e$s that were obtained are displayed in Fig. 5. Here the built-in module in MWS was used and the Balleyguier method in HFSS.

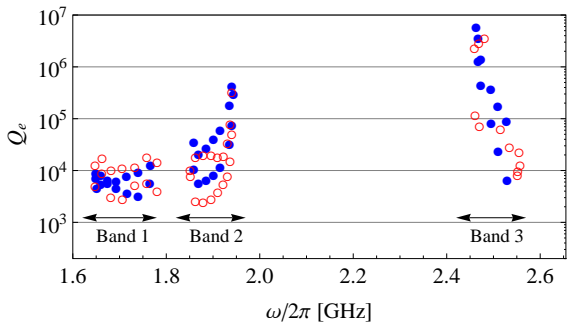

Figure 5: HOM damping Qs results calculated using HFSS (blue dots) and MWS (red circles).

The $Q_e$s of the third band are almost as large as $10^7$. From a beam dynamics perspective these are considerably large. However, modifying the angular separation from 115 degrees to 90 degrees improves the overall damping of the modes. This is illustrated in Fig. 6.

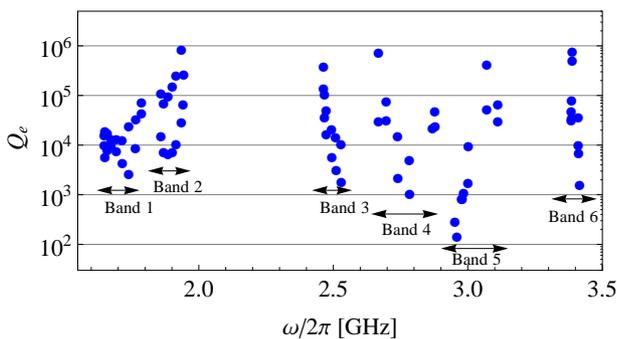

Figure 6: HOM damping Qs results of new HOM couplers configuration calculated using HFSS.

As before, simulations were performed using HFSS and the $Q_e$s were obtained using Balleyguier method. The couplers in this configuration are better targeted to couple to the e.m. fields. The modes are well-damped as the $Q_e$s of the first six dipole bands are below $10^6$.

This new configuration has the potential to be utilised in the NLSF cavity for the ILC.

## CONCLUDING REMARKS

An initial design of both the fundamental power coupler and HOM couplers have been presented. The FP coupler is based on the TTF-III type coupler. Furthermore, the baseline TESLA HOM coupler design is used for the NLSF cavity to damp the HOMs. An accurate calculation of the $Q_e$ using the Balleyguier method has been validated both in MWS and HFSS. Extensive beam dynamics particle tracking simulations are necessary to confirm the validity of the present design. These will indicate whether the HOMs are adequately suppressed to a level which does not appreciably dilute the beam emittance.

## ACKNOWLEDGEMENT

We have benefited from discussions at the MEW group meetings at the Cockcroft Institute. N.J. receives support from the Royal Thai Government and the Thai Synchrotron Light Research Institute [8].

## REFERENCES

[1] N. Juntong and R.M. Jones, High-Gradient SRF Cavity with Minimised Surfaced E.M. Fields and Superior Bandwidth for the ILC, SRF2009, THPPO024, 2009
[2] P. Balleyguier, External Q Study for APT SC-Cavity Couplers, LINAC98, MO4037, 1998.
[3] N. Juntong, Investigation of Optimised Electro-magnetic Fields in SRF Cavities for the ILC, The University of Manchester, PhD thesis, 2011.
[4] The International Linear Collider Reference Design Report, 2007.
[5] www.cst.com/Content/Products/MWS/
[6] www.ansoft.com/products/hf/hfss/
[7] W.-D. Möller for the TESLA collaboration, High Power Coupler for the TESLA Test Facility, 9$^{th}$ SRF Workshop, Santa Fe, Vol. 2, p. 577-581, 1999
[8] www.slri.or.th/en/